\documentclass[onecolumn,showpacs,aps]{revtex4}
\usepackage{color}
\usepackage{mhchem}

\usepackage{graphicx}
\usepackage{amsfonts, amsmath, amstext, amssymb, amsfonts, amsxtra}
\usepackage{braket}
\usepackage{bbm}
\usepackage{bbold}
\usepackage[protrusion=true,expansion=true]{microtype}
\usepackage{wasysym}
\usepackage{cleveref}
\usepackage{color}
\usepackage{ulem}

\begin{document}
	
	\title{Genetic noise mechanism for power-law switching in bacterial flagellar motors}
	
	\author
	{
		M.I.~Krivonosov$^{1}$, V.~Zaburdaev$^{1,3}$, S.V.~Denisov$^2$, and M.V.~Ivanchenko$^2$
	}
	\email
	{
		ivanchenko.mv@gmail.com
	}
	
	\affiliation
	{
		$^{1}$Institute of Supercomputing Technologies, Lobachevsky University, Nizhny Novgorod, Russia \\
		$^{2}$Department of Applied Mathematics, Lobachevsky University, Nizhny Novgorod, Russia \\
		$^{3}$Max Planck Institute for the Physics of Complex Systems, Dresden, Germany \\
		$^{4}$Institut f\"ur Physik, Universit\"at Augsburg, D-86135 Augsburg, Germany
	}

	% Mail !!!
	
	%\keywords {}

	\begin{abstract}
		{Switching of the direction of flagella rotations is the key control mechanism governing the chemotactic activity of \textit{E. coli} and many other bacteria.
		Power-law distributions of switching times are most peculiar because their emergence cannot be  deduced from simple thermodynamic arguments. Recently  it was suggested 
		that by adding finite-time correlations into Gaussian fluctuations regulating the energy height of barrier between the two rotation states, it is possible
		to  generate a switching statistics with an intermediate power law asymptotics. By using a simple model of a regulatory pathway, we demonstrate that the required 
		amount of correlated `noise' can be produced by finite number fluctuations of reacting protein molecules, a condition common to the intracellular chemistry. 
		The corresponding power-law exponent appears as  a tunable characteristic controlled by parameters of the regulatory pathway network such 
		as equilibrium number of molecules, sensitivities, and the characteristic relaxation time. }
		
	\end{abstract}
	
	\maketitle
	
	\section{Introduction}

Bacteria are ubiquitous in nature, displaying a fascinating diversity in size, shape and habitat. 
Often, they cooperate to build colonies, also known as biofilms to adapt to changing and hostile environments \cite{biofilm}. 
Cells employ different strategies of taxis, sensitivity to temperature, chemical or electrical field gradients to 
vary direction of motion, reach favorable niches, and avoid harmful substances \cite{chemotaxis}. 
Among them, chemotaxis is probably best studied and its mechanisms are now understood quite well \cite{bacterialchemotaxis}.
	
	Locomotion of {\it E.~coli} is the one of the most popular case study for bacterial swimming \cite{imaging}. {\it E.~coli} cells have several flagella 
	that can rotate clockwise (CW) or counterclockwise (CCW). When all flagella are in a CCW rotation, they form a bundle and the cell performs a 
	directed motion, often referred to as ``run''. When one or several flagella switch to CW regime, the cell stops and begins to ``tumble'' \cite{imaging}. 
	During this stage {\it E.~coli} chooses the direction of motion for the next run. The resulting angles are randomly distributed with the 
	mean about $70^o$  with respect to the direction of the previous run. 
	A similar motility pattern is also found in a number of marine bacteria, for example {\it S.~putrefaciens}, {\it P.~haloplanktis} or {\it V.~alginolyticus}, 
	although they have a single flagella and the mean turning angle is often close to $180^o$, corresponding to reversals of the direction of motion \cite{tracking}. 
	
Chemotactic strategies rely on controlling frequency of switching between CW and CCW rotations of bacterial motors: 
a favorable signal increases the intervals of directed motion. By regulating the duration of linear motion, bacteria perform 
a random walk biased towards (or away) the source of a chemical attractant (or repellent) \cite{modelingchemotaxis}. In the absence 
of chemical gradients, durations of runs were commonly believed to be exponentially distributed \cite{kineticschemotaxis}. However, recent advances in single cell 
tracking demonstrated strong cell to cell variability {\cite{ecoli3d,corrmotileprotein}}, and, in observations of single motor rotations, even the power-law distributions
were reported \cite{korobkova, modelingchemotaxis}.

The signaling pathway that modulates switching of individual flagellar motor rotation in response to a chemical signaling is well understood by now, 
and the corresponding biophysical models reproduce experimental observations quite successfully \cite{modelingchemotaxis,chemotaxisdiversity}. However, 
the relative complexity of the full model, as well as some uncertainty about the influence of various intra- and extracellular processes, leave the question 
of the origin of the power-law distribution still open, to some extent. Moreover, to the best of our knowledge, power-law distributed runs were not yet observed 
experimentally with freely swimming bacteria \cite{exp}.
 
Considerable progress in understanding the statistics of motor switching was achieved by means of a minimal model considering transitions between the two states 
over an energy barrier \cite{noisechemotaxis}. The regulating pathway was reduced to the action of the phosphorylated form of a signaling molecule, CheY-P, 
such that higher concentration of CheY-P leads to a higher probability of CCW to CW transition \cite{steadystate}. It was found, that Gaussian fluctuations with a 
finite correlation time in the height of these barriers can produce an algebraic scaling in the distribution of CCW durations \cite{steadystate}. These studies 
are in line with a broader research trend aimed at understanding of the genesis of  power-law distributions in  Langevin systems with multiplicative  nosie \cite{m1,m2,m3}.
However, the origin of such fluctuations was not well-established in Ref.~\cite{steadystate}. 
It was conjectured that intrinsic stochasticity of the regulating genetic pathway, in particular, `genetic noise' 
due to a finite number of reacting protein molecules in the cell could produce such fluctuations.

In this paper, we investigate this hypothesis by using chemical kinetics models of CheY-P synthesis and resulting CCW -- CW switching. 
We demonstrate that molecular noise and its correlations due to finite regulator protein synthesis timescales are sufficient to reproduce 
distribution of CCW durations with intermediate power-law asymptotics. Complementary exponential distributions of CW durations are consequence of a weaker sensitivity 
of CW -- CCW transition threshold to CheY-P regulation.

The  paper is organized as follows: In Section II, we formulate two models of CheY-P regulating pathways, specify the corresponding rates, and 
briefly discuss meaning of model parameters. In Section III, we present the results of a massive numerical sampling
obtained by implementing Gillespie stochastic algorithm. An alternative approach, based on the so-called full-counting statistics,
which allows us to avoid resource expensive sampling and extract the probability distribution functions for CCW and CW durations directly
from the chemical rates, is presented in Section IV.  We conclude the paper with some discussion in Section V.

	\section{Models}

	\begin{figure*} [!t]
	  		(a){\includegraphics[width=0.45\columnwidth]{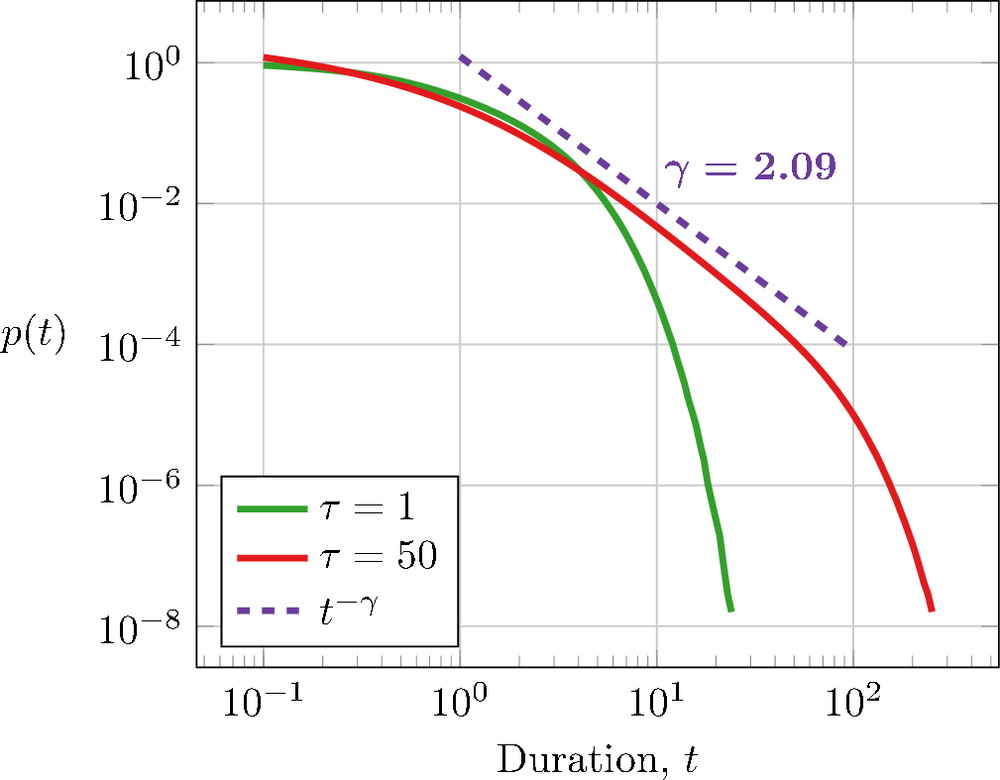}}  	
	  		(b){\includegraphics[width=0.45\columnwidth]{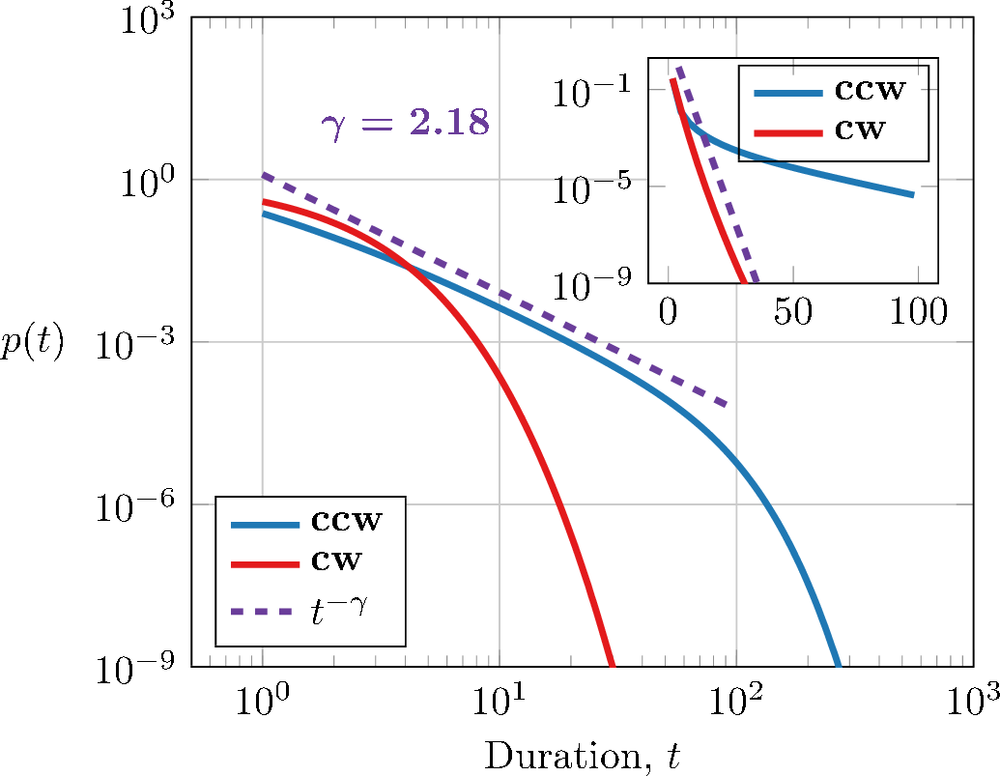}}
	 	\caption
	 	{
	(Color online) (a) Probability density functions of durations in the CCW 
	state  for different values of relaxation timescale of the CheY-P subsystem, $\tau$. 
	The other parameters are $\alpha=10$, $Y_0=20$, and $K_0=1$; (b) 
	Probability density functions of CW and CCW durations for $\alpha^+=1$, $\alpha^-=10$, $Y_0=20$, $\tau=50$, $K_0=1$. 
	Dashed lines are power-laws with exponents $\gamma=2.09$ (a) and $2.18$ (b), which coefficient of determination is $R^2=0.98$, the number of samples is $N_{ccw}=N_{cw}=10^{10}$, $t \in [1; 100]$. 
	Inset: the same functions on a semi-log plot.
		}
	 	
	 	\label{fig:1}
	 	
	 \end{figure*}	
	
Chemotactic regulation of {\it E. coli} relies on the complex intracellular signaling network \cite{korobkova}. It 
starts with chemoreceptors  at cytoplasmic membrane, the binding sites for chemoattractant molecules. The intracellular 
transduction cascade controls production of CheY-P protein that diffuses to motors and modulates switching of the direction of rotation, CW or CCW, making bacteria tumble or run.

	 \begin{figure*} [t!]
	  		(a){\includegraphics[width=0.45\columnwidth]{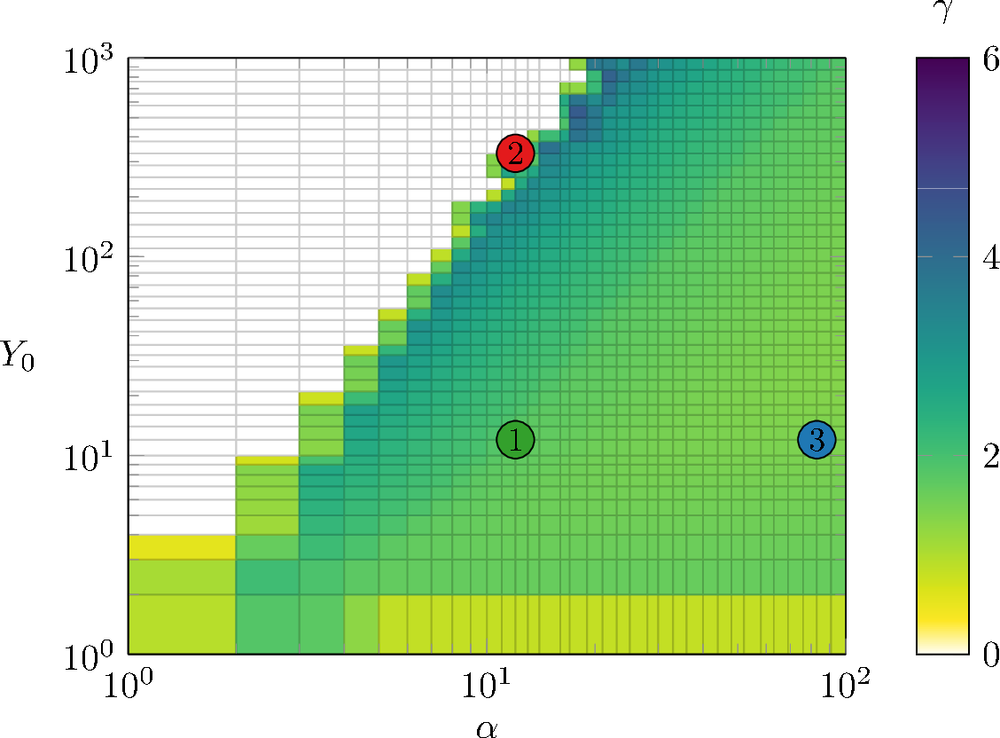}}  
				(b){\includegraphics[width=0.45\columnwidth]{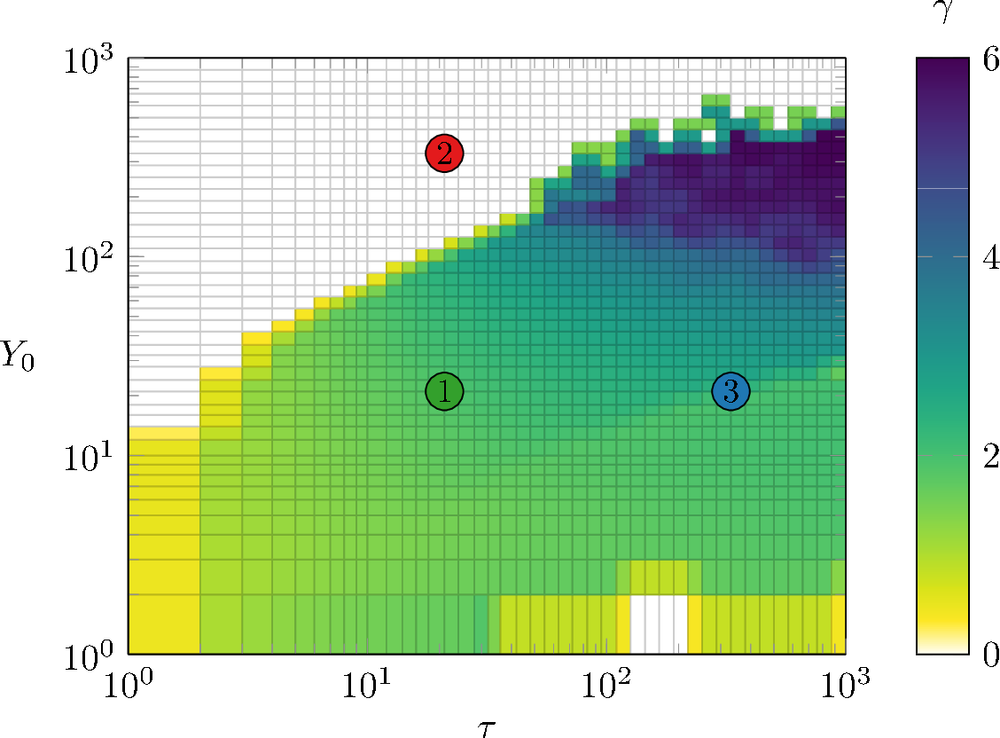}}  
			(c){\includegraphics[width=0.45\columnwidth]{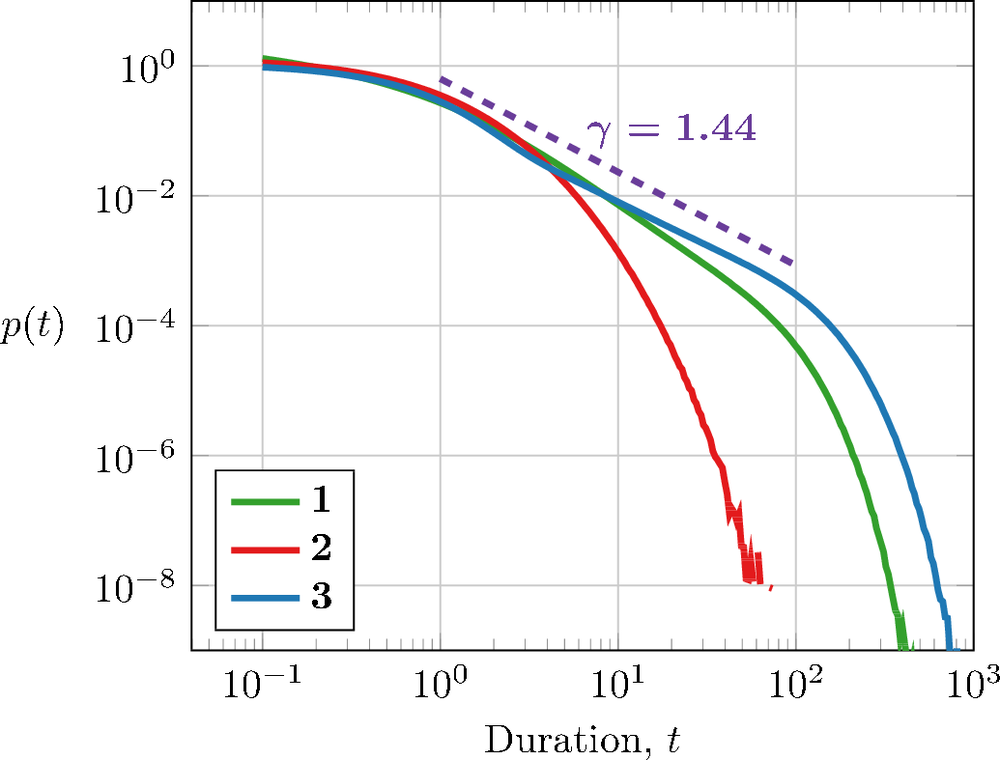}}  
	  				(d){\includegraphics[width=0.45\columnwidth]{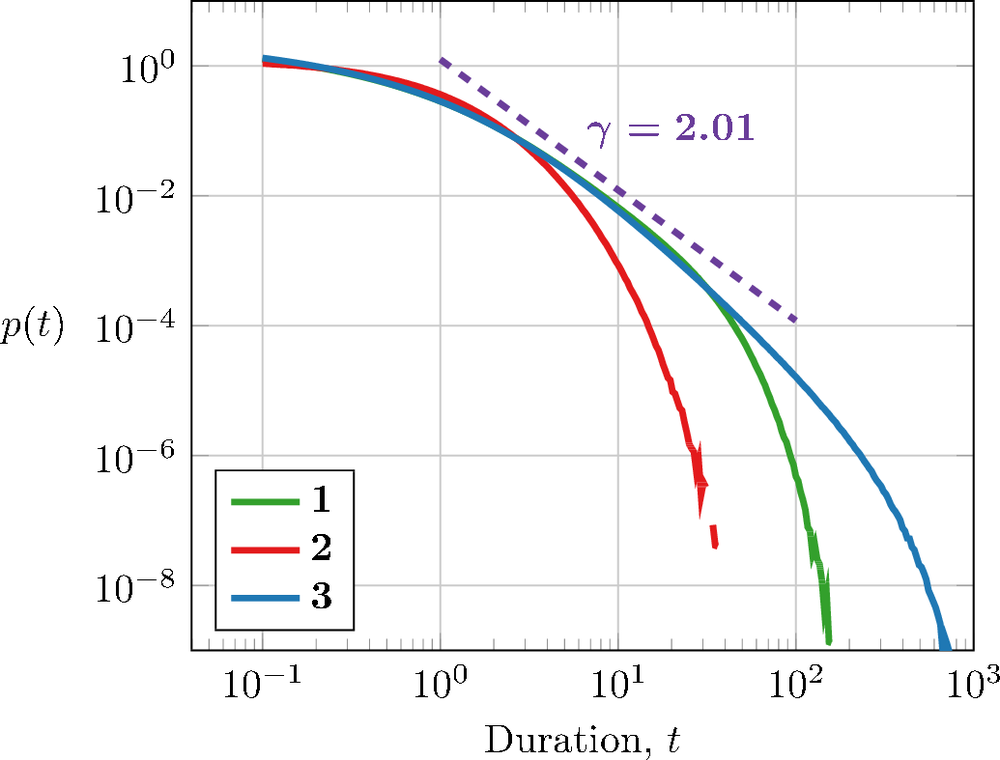}}  
	 	\caption
	 	{
	 		(Color online) Power law exponent $\gamma$ for the probability density functions (PDFs) of CCW durations (color coded) 
	 		as function of the mean number of CheY-P molecules, $Y_0$  and sensitivity $\alpha$ [for fixed $\tau=50$](a)  
	 		or correlation time $\tau$ [for fixed $\alpha=10$ and $K_0=1$] (b). 
	 		In the white area the power law fit for duration distributions does not pass the quality test 
	 		(cf. Models and Methods for details.) PDFs for the points marked on panel (a, b) are shown on panels (c, d), respectively.
	 	}
		\label{fig:2}
	 	\end{figure*}

A minimal model of the regulatory pathway is given in terms of chemical kinetics. Transitions between states with different number of CheY-P molecules, $Y$, follow 
	\begin{equation}
		\label{eq:1}
		\ce{Y <-->[\ce{K_{y}^+}][\ce{K_{y}^-}] {Y+1}},
	\end{equation}
	where $K_{y}^+=\frac{Y_0}{\tau}, K_{y}^-=\frac{Y+1}{\tau}$ are transition rates, $Y_0$ is the equilibrium number of molecules, 
	and $\tau$ is the characteristic relaxation time of the signaling pathway. Being an elementary birth-death process, it intrinsically 
	contains all necessary ingredients, stochasticity, discreteness of states and finite correlation time, $\tau$, which previously had to 
	be brought in by an additive Gaussian noise \cite{noisechemotaxis}. We assume that concentration of chemoattractant, the input signal for the pathway, 
	is changing slowly (in course of motion of a cell in the gradient or time variations of its level), as compared to the switching timescale, so that $Y_0$ can be taken constant. 

Switching of flagella rotations is also modeled from the first principles. Let $X=0$ correspond to the clockwise and $X = 1$ to the counterclockwise regimes. Transitions are controlled by the state of the regulating pathway, $Y$ \cite{noisechemotaxis}
	\begin{equation}
		\label{eq:2}
		\ce{X <-->[\ce{K^+_x}][\ce{K^-_x}] {X+1}},
	\end{equation}
	where $K_x^+=K^+(1-X), K_x^-=K^-X$, formally restricting transitions to the set of two states $X=\{0,1\}$. The transition rates are modulated by the level of CheY-P: 
	\begin{equation}
		\label{eq:3}
		K^{\pm}=K_{0} \exp\left(\pm\alpha^\pm\frac{Y_0-Y}{Y_0}\right),
	\end{equation}
	where $\alpha^\pm>0$ set sensitivities, and the energy barriers are approximated by a linear dependence on CheY-P level \cite{steadystate}. 
	
Let us discuss the qualitative behavior of the model. Assume that a flagellum rotates clockwise (tumbling), such that $X = 0$, then $K_x^+=K^+, K_x^-=0$. 
If the level of CheY-P goes below the equilibrium value, $Y < Y_0$, then the rate coefficient $K^+ > K_0$ and 
switching to counterclockwise rotation (running) will be favoured. Conversely, higher levels of Che-P, $Y > Y_0$, will delay the switch. When a flagellum 
rotates counterclockwise (run), respective fluctuations above and below an equilibrium value $Y_0$ will lead to the opposite effects. Non-identical 
sensitivities to the regulating signal, $\alpha^\pm$, allow for independent tuning of the transition rates between running and tumbling. 

The above simple model has one drawback, that is transition rates in Eq.(\ref{eq:3}) can get exponentially large (small) in response to increasing (decreasing) signal $Y$, 
which may be not biologically plausible. To account for a finite capacity of CheY-P -- motor protein binding, the rates can be modified as suggested in Ref.~\cite{chemotaxisdiversity}:
	\begin{equation}
		\label{eq:4}
		K^{\pm}=K_0^{\pm} \exp\left[ \pm\frac{\alpha^\pm}{2}\left(\frac{1}{2}-\frac{Y}{Y+K_d}\right)\right],
	\end{equation}
		which corresponds to saturation with the level of $Y_n$ above the dissociation constant $K_d$.

	%===========================================================================		

	\section{Results of the stochastic sampling}
	
We first perform numerical simulations of the models (\ref{eq:1}-\ref{eq:4}) by implementing Gillespie stochastic algorithm \cite{gillespie}, 
thereby making the study free of any kind of approximation in terms of deterministic mean-field equations or  continuous-state (Langevin) approximations \cite{chemotaxisdiversity,noisechemotaxis}.
%, also taking into account integer molecule numbers. 
Each realization corresponds to $N = 10^7$ steps of the algorithm 
(one of a possible set of chemical reactions). The durations of residence in each state, $t_{ccw}$ and $t_{cw}$, are determined as the time between two consecutive relevant
reactions. At least $N_{cw}=N_{ccw}=10^{10}$ durations of CW and CCW states are collected in repeated realizations to calculate their probability distribution functions (PDFs), 
$p(t_{cw})$ and $p(t_{ccw})$, and analyze their dependence on the relaxation (correlation) time of the regulatory signal and the sensitivity of transition rates. 
	
The power-law fitting of the obtained PDFs is then performed by using the least squares linear regression for the log-log scaled distributions. 
The quality of fit is characterized by the coefficient of determination \cite{R2}, $R^2\in[0,1]$, with large values corresponding to better fit. 
We vary the interval of durations to search for the best fit, $l$, and request that in spans over at least $1.3$ decades with $R^2>0.98$. Otherwise, the power law hypothesis is rejected.

We start with the simplest form of rate coefficients, as given by Eq.~(\ref{eq:3}), and estimate PDFs of durations of run and tumble phases. 
It is straightforward to see that if CheY-P is absent, $Y_0=0$ (switching is 
insensitive to CheY-P, $\alpha^\pm=0$), the process is Poissonian and the PDFs of residence intervals are exponential, $p(t_{ccw}), p(t_{cw})\propto \exp(-K_0t)$. 

We begin with the case of equal sensitivities, $\alpha^\pm=\alpha$. 
%Although, strictly speaking, the rate coefficients for transitions in the opposite directions are not identical, we have not found any substantial difference in the resulting distributions and further make use of the one for CCW state $p(t_{ccw})$, for illustration. 
Numerical results indicate that a correlated molecular noise due to 
the pathway signal, $Y$, can produce an algebraic scaling in distributions of 
durations, $p(t_{ccw})\propto t^{-\gamma}$, over a pronounced interval ($1.5-2$ decades) 
with the cutoff at large durations; see Fig.~ \ref{fig:1}(a). PDFs with intermediate power-law asymptotics extending over $\lesssim 2$ decades 
and then damped with a near exponential cut-off are typical to both \textit{in vivo} and \textit{in vitro} measurements of bacteria and cell spatial activities; 
see, e.g.,  Refs.~\cite{korobkova,haris}.

In accordance with Ref.~\cite{noisechemotaxis}, 
decreasing the correlation time changes this distribution towards an exponential. 
The model is also capable of reproducing the exponential statistics for 
CW durations, coexisting with the power law scaling for CCW durations. 
The only source of non-identity is in the sensitivity of transition rates to signaling, $\alpha^+\neq\alpha^-$. 
As the residence time in a state is determined by the escape rate, it can be expected that a decreased sensitivity 
of CW $\rightarrow$ CCW transition to the level of CheY-P, will lead to an exponential distribution of CW durations 
(as in the limiting case $\alpha^+=0$). An example of this regime is shown in Fig.~\ref{fig:1}(b).

	 \begin{figure} [b!]
	 	
	  		{\includegraphics[width=0.45\columnwidth]{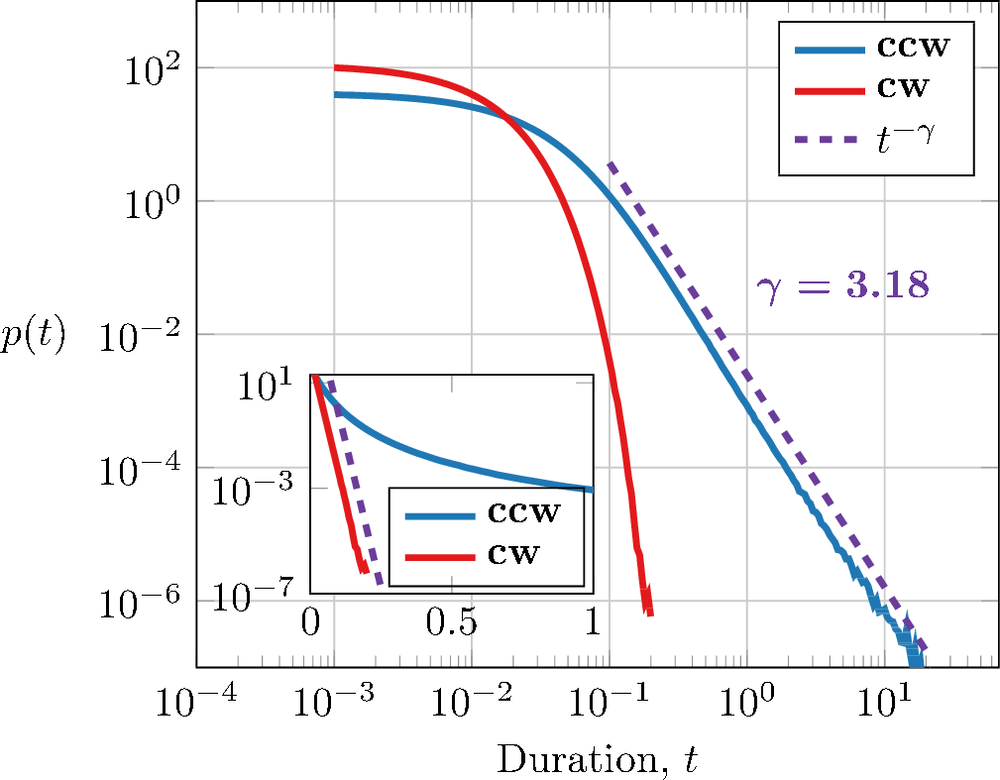}}  
	 	\caption
	 	{
		(Color online) Probability density functions  of CW and CCW durations 
for $\alpha^-=30, \alpha^+=1$, $Y_0=10$, $\tau=50$, $K_d=2$, $K^+_0 = 100$, $K^-_0 = 0.2$. 
Dashed line is a  power-law with $\gamma=3.18$, $R^2=0.999$, $N_{ccw}=10^{11}$, $t \in [0.1; 20]$. 
Inset: the same distributions in the semi-log plot.}	 	
	 	\label{fig:3}
	 	
	 \end{figure}

Extensive simulations in a large parameter region  reveal  that  power law distributions  
emerge only when the relaxation timescale of CheY-P level is substantially greater then that of switching between CW and CCW states, $\tau\gg1/K_0$, see Fig.~\ref{fig:2}. 
At the same time, we observe that the increase of the mean number of signaling molecules, $Y_0$, destroys power law scaling, see Fig.~\ref{fig:2}(a).
This effect can be understood as diminishing of the fluctuations with large numbers of reacting molecules. 
Sensitivity parameter, $\alpha$, is also able to control transition between power law and exponential PDFs of duration intervals (Fig.~\ref{fig:2}(b)). 
Values of the power-law exponent found in most of the region, $1<\gamma<2$, are consistent with experimental observations, 
which estimated the power law exponent for the cumulative distribution of CCW durations as $\gamma-1\approx1.5$, and evidence that slow 
methylation as a part of the signaling pathway, is responsible for the long time correlations in the signal output that could stand behind the power law \cite{korobkova}.

	 \begin{figure*} [t!]
	 	 		(a){\includegraphics[width=0.45\columnwidth]{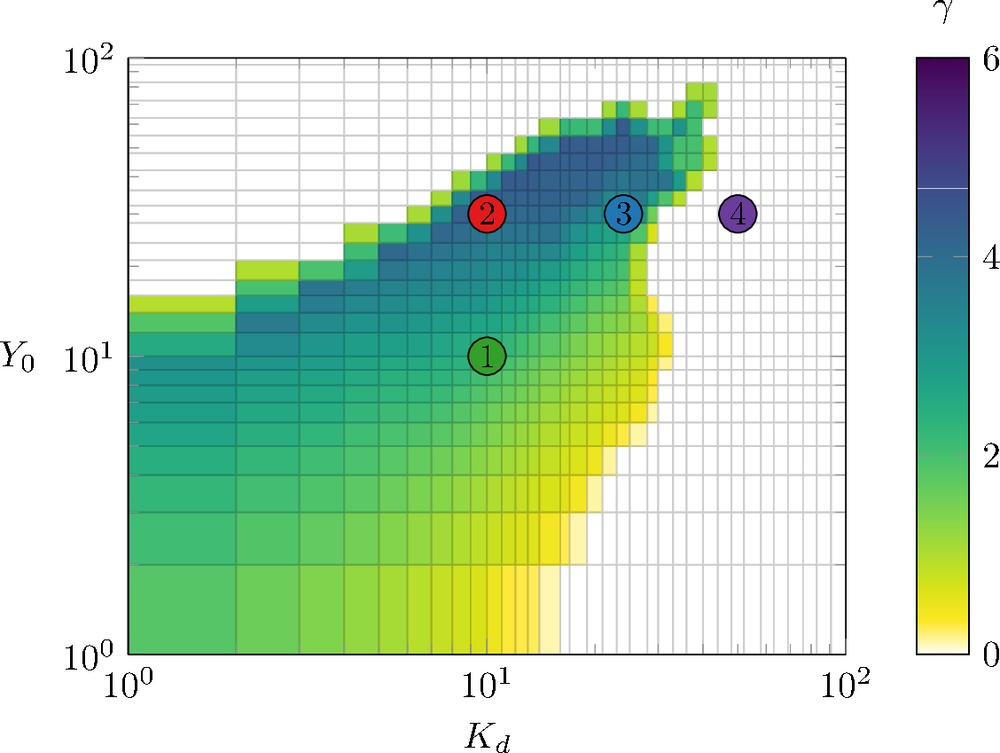}}  
				(b){\includegraphics[width=0.45\columnwidth]{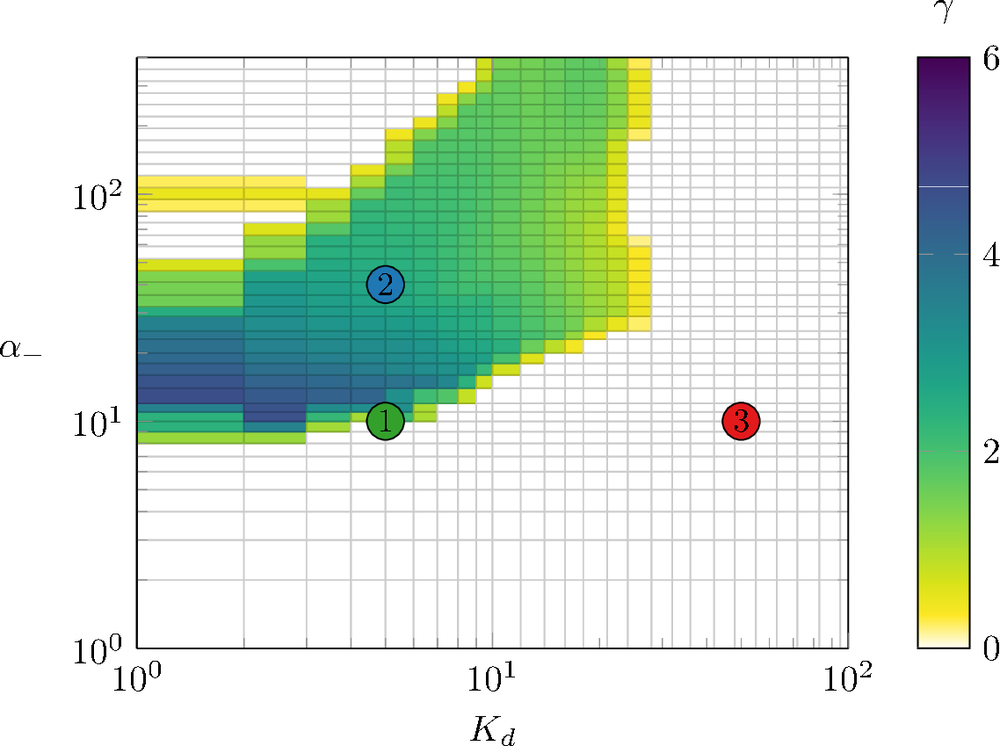}}  
	 	 		(c){\includegraphics[width=0.45\columnwidth]{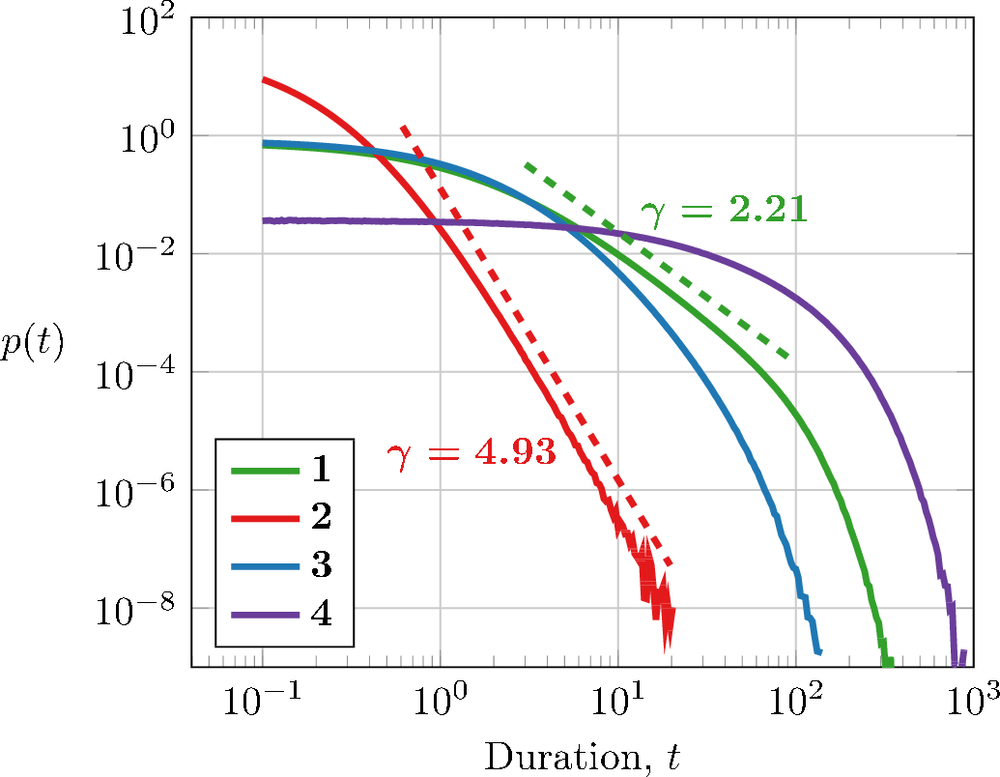}}  
				(d){\includegraphics[width=0.45\columnwidth]{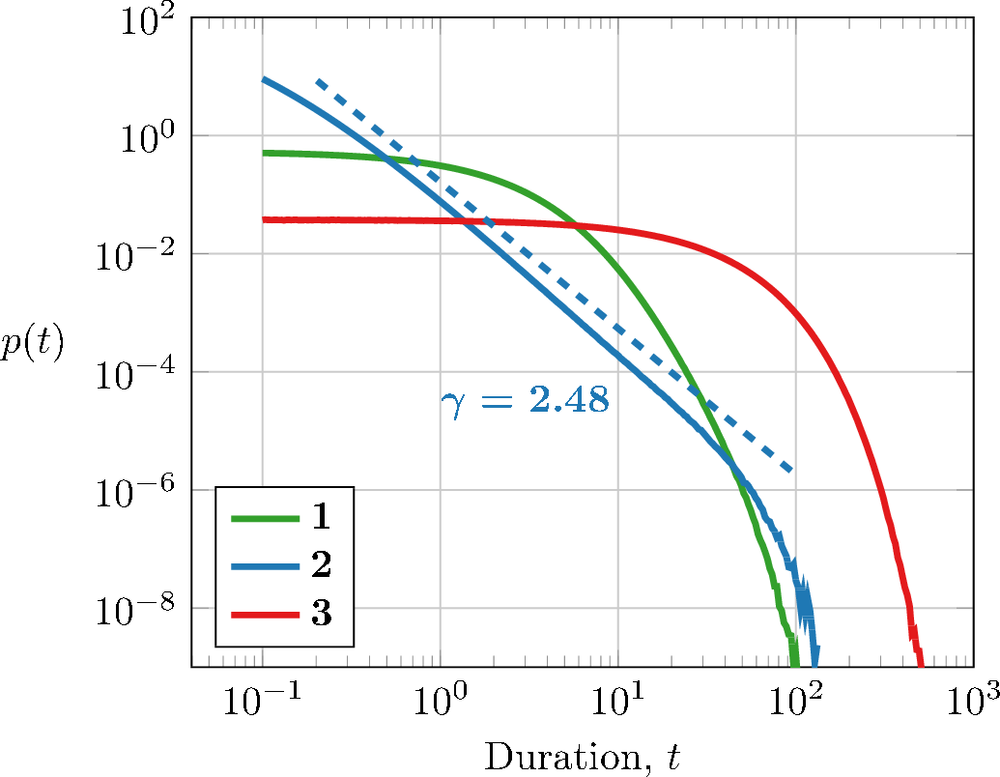}}
	 	\caption
	 	{
	 		(Color online) Power law exponent $\gamma$ estimated for the PDFs of  CCW durations (color coded) as function of
	 		saturation constant, $K_d$, and (a) 
	 		mean number of CheY-P molecules, $Y_0$, for fixed $\alpha^-=30$ and (b) sensitivity $\alpha^{-}$ for fixed $Y_0=10$. 
	 		The other parameters are $\alpha_+=1$, $\tau=50$, $K_{+0} = 100$, $K_{-0} = 0.2$. In white area, the power law fit for 
	 		the PDFs does not pass the quality test (cf. Models and Methods for details.)
	 		PDFs for the points marked on panel (a, b) are shown on panels (c, d), respectively.
	 	}
		\label{fig:4}
	 	\end{figure*}

	Now we consider a more biologically plausible model for transition rates, Eq.~(\ref{eq:4}), that captures saturation in the 
	motor response. To sharpen the effect, we turn to the case $K_d < Y_0$, when saturation is pronounced already on the equilibrium level of CheY-P, $Y_p=Y_0$. 
	Our results confirm that again slow relaxation and long correlations in $Y_p$, together with higher sensitivity of the CCW$\rightarrow$CW transition to the regulatory signal, 
	can lead to the appearance of a power law asymptotics in the PDF of CCW durations, while CW durations remain exponentially distributed, see Fig.\ref{fig:3}. 
	
A systematic study of statistics as a function of the parameter values of the model is presented in Fig.~\ref{fig:4}. 
The $(K_d,Y_0)$ parameter plane exhibits two different regions, Fig.~\ref{fig:4}(a). Namely, for $Y_0<K_d$, when saturation effects are weak, 
there is a strong deviation from the power-law statistics as the number of molecules increases.  Power law scaling appears with relatively low numbers, $\gamma<2$. 
For $Y_0>K_d$ saturation effects make the exponent to reach larger values, $\gamma>4$.   

On other parameter plane, $(K_d,\alpha^-)$, we observe a persistence of the power-law scaling in a large range of CCW$\rightarrow$CW transition sensitivities, $\alpha^{-}$, Fig.\ref{fig:4}(b). 
There, one again notices the regimes with relatively small, $\gamma\leq2$, and large, $\gamma>4$, values of  the power law exponent, as dictated by the ratio between the mean number 
of CheY-P molecules, $Y_0$, and the saturation constant $K_d$.

	%===========================================================================	
	
	 \begin{figure*} [t!]
	 	 		(a){\includegraphics[width=0.15\columnwidth]{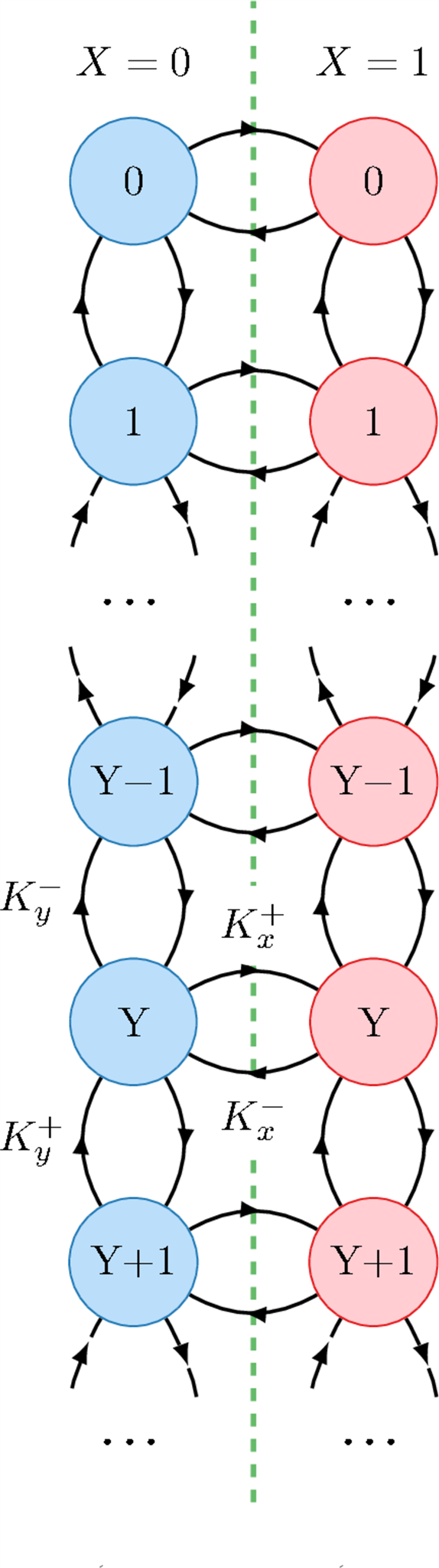}}  
				(b){\includegraphics[width=0.45\columnwidth]{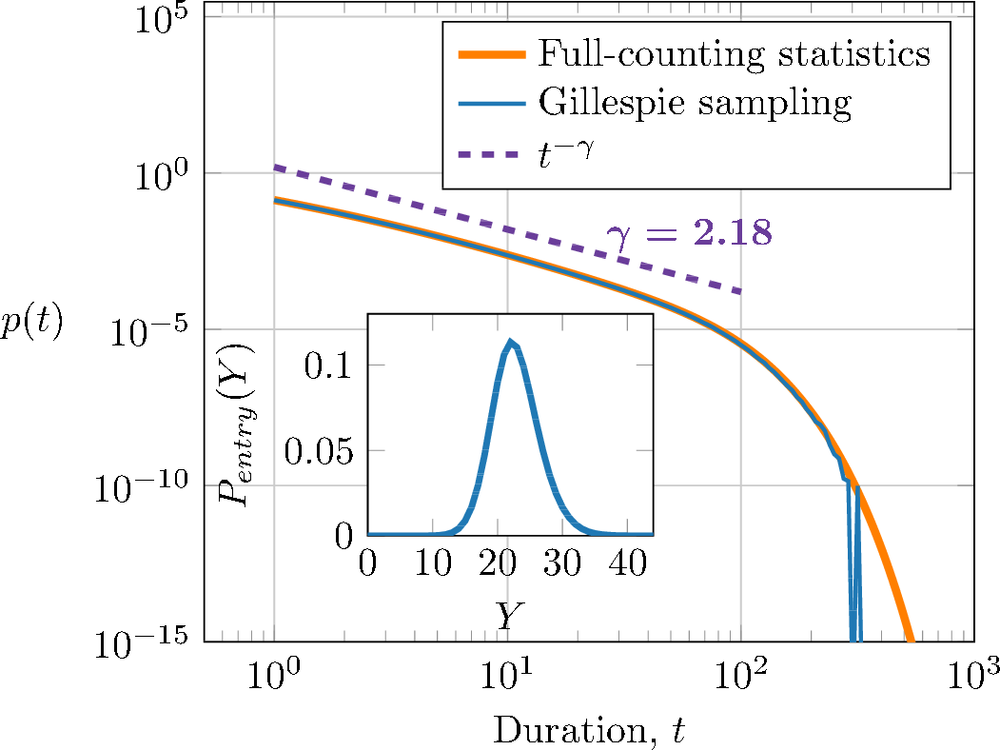}}  
	 			
	 	\caption
	 	{
	 		(Color online) (a) Transition diagram for the process (1-2). (b)  Probability density functions  of CCW durations
	 		obtained from  full-counting statistics of the process described by the master equation Eq.~(\ref{eq:Master}) with truncation at $Y_{max}=100$
	 		(thick orange line) compared to the results
	 		of Gillespie sampling (blue line). The inset shows the probability distribution $P_{entry}(Y)$ for the entry distribution of the CCW phase, Eq.~(\ref{eq:entry}). 
	 		Other parameters are the same as in Fig.~1b.
	 	}
		\label{fig:5}
	 	\end{figure*}

\section{Switching time distributions from full-counting  statistics}

Costly numerical sampling can be avoided by implementing  the toolbox of full-counting statistics \cite{esposito}. Originally introduced in the context of quantum transport \cite{lesovik},
this formalism  applies to all master equations, independent of their genesis \cite{emary}. Here we demonstrate how it  can be used to extract switching statistics
directly from transition rates $K_x^{\pm}$(Y) and $K_y^{\pm}(Y)$.

The kinetics of the models (1 - 2) can be evaluated in terms of continuous-time Markov process \cite{markov}; the corresponding transition diagram is shown in Fig.~5a. 
A state $Z(t)$ of the process at time $t$ is a two-dimensional vector, $Z = \{X,Y\}$  with first binary component representing the rotational regime, $X=0$ (CW) and $X=1$ (CCW), 
and the second one, $Y$, being the number of CheY-P molecules. The corresponding master equation for the probability vector $P(Z,t) \equiv P(X,Y;t)$ has the form

\begin{eqnarray} 
 \dot{P}(0,0;t)= -\big[\frac{Y_0}{\tau} + K_x^+(0)\big]P(0,0;t) + K^-_x(0)P(1,0;t) + \frac{1}{\tau}P(0,1;t) \\ \nonumber
 \dot{P}(1,0;t)= -\big[\frac{Y_0}{\tau} + K_x^-(0)\big]P(1,0;t) + K^+_x(0)P(0,0;t) + \frac{1}{\tau}P(1,1;t) \\ \nonumber
 ... \\ \nonumber
 ... \\ \nonumber
 ... \\ \nonumber
\dot{P}(0,Y;t)= -\big[\frac{Y_0+Y}{\tau} + K_x^+(Y)\big]P(0,Y;t) + K^-_x(Y)P(1,Y;t) + \frac{Y+1}{\tau}P(0,Y+1;t) + \frac{Y_0}{\tau}P(0,Y-1;t)\\ \nonumber
\dot{P}(1,Y;t)= -\big[\frac{Y_0+Y}{\tau} + K_x^-(Y)\big]P(1,Y;t) + K^+_x(Y)P(0,Y;t) + \frac{Y+1}{\tau}P(1,Y+1;t) + \frac{Y_0}{\tau}P(1,Y-1;t)\\ \nonumber
... 
~~~~~~~~~~~
 \label{eq:Master}
\end{eqnarray}

It is noteworthy that our model serves a generalization of a fluid queue driven by a  birth-death process proposed by van Doorn, Jagers, and  de Wit  \cite{doorn}. 
Namely,  in the framework of Markov processes, reaction (2) is an exchange of probability between two states; it can be considered as an exchange of a fixed amount of an incompressible fluid 
between two tanks
at rates determined by the state of a  birth-death process (in the original formulation there was only one infinite-volume tank and unlimited amount of fluid). 
A  CCW duration is a time a single (marked)  molecule of fluid spends in the tank ``$1$'' before leaving it. By introducing a probability 
vector $\mathbf{P}(t) = \{P(X=0,Y=0;t), P(X=0,Y=1;t), P(X=0,Y=2;t),..., P(X=1,Y=0;t),P(X=1,Y=1;t), P(X=1,Y=2;t),...\}$ 
and  a transition rate matrix $\mathbb{Q}$ \cite{markov}, we can re-write equation (\ref{eq:Master}) in a compact form, $\dot{\mathbf{P}}(t) = \mathbb{Q}\mathbf{P}(t)$.
Matrix $\mathbb{Q}$ consists of two identical semi-infinite diagonal blocks 
 $\mathbb{\tilde{Q}}_{\mathrm{BD}}$ [matrices of the tridiagonal form and consisting of the rates of the birth-death process] accompanied  with  off-diagonal blocks
$-\mathbb{\tilde{Q}}_{\mathrm{01 (10)}}$ [matrices of the diagonal form, with entries $K_x^+(Y)$ ($K_x^-(Y))$]. 

Next we consider the probability $\mathcal{P}(t)$ of no transition occurred from state $1$ to state $0$ during time $t$ . This probability is specific to the initial 
distribution $\mathbf{P}_{entry} \equiv \{P_{entry}(Y)\}$, $P_{entry}(Y) := P(1,Y,0)$ (the meaning of this notation will be apparent soon). 
For a chosen initial state, $\mathcal{P}(t)$ can be calculated  from the state vector $\mathbf{P}^{(0)}(t) \equiv \{P^{(0)}(Y;t)\}$ (here superscript $(0)$ denotes
number of transitions on which the state is conditioned \cite{emary}). 
%of the $X=1$ sub-system at a time $t$ conditioned on that no transition has occurred prior to this time. 
The evolution of this vector is governed by the master equation
\begin{eqnarray} 
 \dot{\mathbf{P}}^{(0)}(t)= \mathbb{\tilde{Q}}_{\mathrm{NT}}\mathbf{P}^{(0)}(t) =  \big[\mathbb{\tilde{Q}}_{\mathrm{BD}} - \mathbb{\tilde{Q}}_{\mathrm{10}}\big] \mathbf{P}^{(0)}(t).
 \label{eq:Master2}
\end{eqnarray}
The no-transition probability is then $\mathcal{P}(t) = \sum_{Y} P^{(0)}(Y;t)$.

Matrix $\mathbb{\tilde{Q}}_{\mathrm{NT}}$ is a semi-infinite Jacobi matrix, i.e., it is a tridiagonal matrix with positive 
off-diagonal entries. Therefore, it has a purely real non-degenerated  spectrum, $\{\lambda_1, \lambda_2,...,\lambda_i,...\}$. The largest eigenvalue is negative, $\lambda_1 < 0$ [there is a 'leak' 
from state $X=1$ to state $X=0$ so that master equation (\ref{eq:Master}) does not preserve the vector norm] and, therefore, $\forall \lambda_i < 0$. 
The conditioned probability vector at any instant of time can be obtained by implementing spectral decomposition, 
%\begin{eqnarray} 
$\mathbf{P}^{(0)}(t) = \sum_i \alpha_i \exp(\lambda_it)\cdot\mathbf{v}_i$,
% \label{eq:decomp}
%\end{eqnarray}
where $\alpha_i = \mathbf{w}^T_i \cdot \mathbf{P}_{entry}$ and $\{\mathbf{w}_i,  \mathbf{v}_i\}$ is the dual left-right eigenset of $\mathbb{\tilde{Q}}_{\mathrm{NT}}$.
The decay of $\mathcal{P}(t)$ is monotonous due to the properties of the spectrum.

Finally, to calculate $p(t)$ for the CCW phase, we need to calculate $\mathcal{P}(t)$ for the initial vector $\mathbf{P}_{entry}$ corresponding to the distribution over $Y$ states 
at the moment when the transition $X=0 \rightarrow X=1$ happens. In this case $p(t) = - \dot{\mathcal{P}}(t) = \sum_i \alpha_i\cdot(-\lambda_i)\cdot\exp(\lambda_it)\cdot\mathbf{v}_i$. 

The entry distribution, in its turn, can be obtained from the stationary distribution  $\mathbf{P}_{st}$ of the full Markov 
evolution, $\mathbb{Q}\mathbf{P}_{st} = \mathbf{0}$, by taking only $X=0$ sub-vector, $\{P_{st}(X=0,Y)\}$, which defines the probability
to find the system in a state $Y$ conditioned on that the system is in the set  $X=0$. Next we should multiply this probability with the probability
that the reaction which takes the system into the set $X=1$ will happen when the system is in the state $\{X=0,Y\}$. This probability is proportional to $K_x^+(Y)$ \cite{problem}, so  we have
\begin{eqnarray} 
P_{entry}(Y) = P_{st}(X=0,Y)\cdot K_x^+(Y)/\mathrm{Norm}[P_{st}(X=0,Y)\cdot K_x^+(Y)],
 \label{eq:entry}
\end{eqnarray}

Practically, it means that all needed ingredients -- (i) the spectrum and eigenset of $\mathbb{\tilde{Q}}_{\mathrm{NT}}$ and (ii) the stationary distribution of $\mathbb{Q}$ -- can be obtained by truncating
the state space with some $Y_{max}$ and performing two routine matrix diagonalizations, of the size $Y_{max}$ and $2Y_{max}$, respectively. A realization of this recipe is presented in Fig. 5b. 
Thus calculated distribution  $p(t)$ perfectly matches the distribution obtained by performing Gillesipe sampling.

	\section{Conclusions}

We studied statistics of flagella switching between CCW (`run') and CW (`tumble') regimes, the output of chemotactic signaling pathway controlled by the molecular dynamics of CheY-P protein. 
We demonstrated that the correlated noise  induced by finite-size fluctuations is sufficient to  reproduce  distributions of CCW durations with intermediate power-law asymptotics 
in a broad parameter region. 
%Exponential distributions of CW durations appear  due to a weaker sensitivity of the CW -- CCW transition threshold to the CheY-P regulation. 
Extensive numerical sampling  allowed us to map  parameter regions with power law statistics and define the values of the corresponding exponents.

While the precise mechanism behind the observed power-law asymptotics needs to be further investigated, their origin can be intuitively understood 
from the Markov interpretation; see Fig.~5a. Namely, it is the presence of hidden variable $Y$ which supervises the transition rates between the two $X$ states (otherwise, in the absence of 
$Y$, $p(t)$ would be a plain exponential distribution). We conjecture that this mechanism is somehow related to the recent deduction of Zip's law for statistical 
models  with hidden ('unobservable', 'latent', etc)
variables \cite{schwab}. Hidden variables 'mix' together many different reactions (processes, pathways, etc) that individually do not obey Zip's law in a such way that the resulting mixture
obeys this law. This mechanism is an interesting  alternative to the prevailing (at the moment) approach based on the fine parameter tuning into criticality. Zipf's law, 
roughly speaking, corresponds to the case with exponent $\gamma = 1$. How this new approach can be generalized --within the chemical kinetics framework -- to the case with tunable 
exponents is a challenging question.

Our results open certain theoretical perspectives. The origin of power-law distributions 
found in mobility patterns of many living organisms, ranging from bacteria to sharks and human beings, 
remains a mystery \cite{levywalks}. Even though one can accept the hypothesis that this type of distributions was 
selected by evolution as the optimal strategy for survival and best accomplishment of every-day routines, 
physiological mechanisms behind the power-laws are not understood yet. A simple chemical network, 
driven by finite number fluctuations intrinsic to intracellular molecular dynamics, is able to generate tunable power-law distributions and therefore constitutes a promising candidate 
for such a 'generator'. Our findings are also of potential relevance to 
bioengineered cell chemotaxis (to control motility of bacteria), biofilms formation and targeted cell-assisted drug delivery.

\section{Acknowledgments} This work was supported by the Russian Science Foundation grant No.~16-12-10496 (M.K., S.D. and V.Z.). 
%and Ministry of Education and Science of the Russian Federation Research Assignment No. 1.5586.2017/BY (M.I.).
	
	%===========================================================================


\begin{thebibliography}{15}
\bibitem{exp} Most of the collected experimental data are in agreement with the exponential run time distribution. There are several models which exhibit this type of run 
		time distributions; see e.g., Refs.~\cite{z1,z2}.
\end{thebibliography}
\end{document}